\documentstyle[aps,preprint,12pt]{revtex}

\widetext
\draft
\tighten
\begin{document}

\title{\normalsize{\rm{\bf HELICITY BASIS AND PARITY}}\thanks{Presented
at the Plebanski Conference ``Topics in Mathematical Physics, General
Relativity and Cosmology", September 17-20, 2002, CINVESTAV, D. F.,
M\'exico and the Jornadas de Investigaci\'on UAZ-2002, Zacatecas, Oct.
8-11, 2002.}}

\author{{\bf Valeri V. Dvoeglazov}}

\address{{\rm Universidad de Zacatecas, Apartado Postal 636,
Suc. UAZ\\Zacatecas 98062, Zac., M\'exico\\
E-mail: valeri@ahobon.reduaz.mx\\
URL: http://ahobon.reduaz.mx/\~\,valeri/valeri.html}}

\date{September 16, 2002}

\maketitle

\bigskip

\begin{abstract}
{\bf Abstract.} We study the theory of the $(1/2,0)\oplus (0,1/2)$
representation in  helicity basis. Helicity eigenstates
are {\it not} the parity eigenstates. This is in accordance with the
consideration of Berestetski\u{\i}, Lifshitz and Pitaevski\u{\i}. 
Relations to the Gelfand-Tsetlin-Sokolik-type quantum field theory are
discussed. Finally, a new form of the parity operator is proposed.
It commutes with the Hamiltonian.
\end{abstract}

\bigskip

First of all, I would like to congratulate Professor J. Plebanski with his
75th birthday. Thank you for your hard work in theoretical physics,
which we all admire.

What are scientific motivations for my talk?
Recently we generalized the Dirac formalism~\cite{Barut,DasG,Dv2,Dv2a} and
the Bargmann-Wigner formalism~\cite{Dv3,Dv5}, and on this basis we proposed a set
of twelve equations for antisymmetric tensor (AST) field; some of them may lead
to  parity-violating transitions. In this paper we are going to study
somewhat related matter, the transformation from the standard basis
to the helicity basis in the Dirac theory. The spin basis rotation
{\it changes} the properties of corresponding states with respect to
parity. The parity is a physical quantum number; so, we try to extract
corresponding physical contents from considerations of the various spin
bases.

Briefly, I repeat the results of ref.~\cite{Dv4,Dv5}. One can find
solutions of the $2(2J+1)$-theory with different parity
properties~\cite{Dv4}.  They can be related to the polarization vectors
obtained by Ruck and Greiner~\cite{Grei}, who found the helicity states
of the 4-vector potential on the basis of the Jackob and Wick
paper~\cite{JW}. Next, I used the generalized Bargmann-Wigner formalism
based on the equations\footnote{The parity-violating
Dirac equation has been derived in~\cite{Dv2a}.
The method of the derivation refers to the van der Waerden, Sakurai and Gersten
works, see references in the previous papers of mine.}
\begin{mathletters} \begin{eqnarray} \left [ i\gamma_\mu
\partial_\mu + \epsilon_1 m_1 +\epsilon_2 m_2 \gamma_5 \right
]_{\alpha\beta} \Psi_{\beta\gamma} &=&0\,,\\ \left [ i\gamma_\mu
\partial_\mu + \epsilon_3 m_1 +\epsilon_4 m_2 \gamma_5 \right ]_{\alpha\beta}
\Psi_{\gamma\beta} &=&0\,,
\end{eqnarray}
\end{mathletters}
Different equations for the antisymmetric tensor field follow
from this set by means of the standard procedure~\cite{Lurie}.
We concluded in~\cite{Dv5} in  part that: 1)
in the $(1/2,0)\oplus (0,1/2)$ representation it is possible to introduce
the {\it parity-violating} frameworks; 2)
the mappings between the Weinberg-Tucker-Hammer formalism for
$J=1$ and the AST fields of the 2nd rank of, at least, eight types exist;
Four of them include both $F_{\mu\nu}$ and $\tilde F_{\mu\nu}$, which
tells us that the parity violation may occur during  the study of the
corresponding dynamics; 3) if we want to take into account the $J=1$
solutions with different parity properties, the Bargmann-Wigner (BW) formalism
is to be generalized; 4) the 4-potentials and the fields in the helicity
basis can be constructed; they have different parity properties comparing
with the standard (``parity") basis; 5) generalizing
the BW formalism in such a way, twelve equations for the AST
fields have been obtained; 6) finally, a hypothesis was
proposed therein that the obtained results are related to the spin basis
rotations and to the choice of normalization.

Beginning the consideration of the helicity basis, we observe that
it is well known that the operator $\hat {\bf S}_3 = {\bbox
\sigma}_3/2\otimes I_2$ does not commute with the Dirac Hamiltonian unless
the 3-momentum is aligned along with the third axis and the plane-wave
expansion is used:
\begin{equation}
[\hat{\cal H},\hat {\bf S}_3]_- = (\gamma^0{\bbox\gamma}^k \times
{\bf\nabla}_i )_3
\end{equation}
Moreover, Berestetski\u{\i}, Lifshitz and
Pitaevski\u{\i} wrote~\cite{Lan}: ``... the orbital angular momentum ${\bf
l}$ and the spin ${\bf s}$ of a moving particle are not separately
conserved. Only the total angular momentum ${\bf j}= {\bf l}+{\bf s}$ is
conserved. The component of the spin in any fixed direction (taken as
$z$-axis is therefore also not conserved, and cannot be used to enumerate
the polarization (spin) states of moving particle." The similar conclusion has
been given by Novozhilov in his book~\cite{Novozh}. On the other hand, the helicity operator ${\bbox\sigma}\cdot
\widehat {\bf p}/2 \otimes I$, $\widehat {\bf p} = {\bf p}/\vert {\bf
p}\vert$, commutes with the Hamiltonian (more precisely, the commutator is
equal to zero when acting the one-particle plane-wave solutions).

So, it is a bit surprising, why the 4-spinors have been studied so well
when the basis was chosen in such a way that they are eigenstates
of the $\hat {\bf S}_3$ operator:
\begin{eqnarray}
u_{{1\over 2},{1\over 2}} = N_{1\over 2}^+\pmatrix{1\cr0\cr1\cr0\cr}\,,
u_{{1\over 2},-{1\over 2}} =N_{-{1\over 2}}^+ \pmatrix{0\cr1\cr0\cr1\cr}\,,
v_{{1\over 2},{1\over 2}} = N_{1\over 2}^-\pmatrix{1\cr0\cr-1\cr0\cr}\,,
v_{{1\over 2},-{1\over 2}} =N_{-{1\over 2}}^-
\pmatrix{0\cr1\cr0\cr-1\cr}\,,\label{sb1}
\end{eqnarray}
and, oppositely, the
helicity basis case has not been studied almost at all
(see, however, refs.~\cite{Novozh,JW}.  Let me remind
that the boosted 4-spinors in the `common-used' basis are
\begin{mathletters}
\begin{eqnarray}
u_{{1\over 2},{1\over 2}} = {N_{1\over
2}^+\over \sqrt{2m (E+m)}}
\pmatrix{p^++m\cr p_r\cr p^- +m\cr -p_r\cr}\,,
u_{{1\over 2},-{1\over 2}} ={N_{-{1\over 2}}^+
\over \sqrt{2m (E+m)}}\pmatrix{p_l\cr p^-+m\cr -p_l\cr p^++m\cr}\,, \\
v_{{1\over 2},{1\over 2}} = {N_{1\over
2}^- \over \sqrt{2m (E+m)}}\pmatrix{p^++m\cr p_r\cr -p^- -m\cr p_r\cr}\,,
v_{{1\over 2},-{1\over 2}} ={N_{-{1\over
2}}^- \over \sqrt{2m (E+m)}}\pmatrix{p_l\cr p^-+m\cr p_l\cr -p^+-m\cr}\,,
\, ,
\end{eqnarray} \end{mathletters}
$p^\pm = E\pm p_z$, $p_{r,l}= p_x\pm ip_y$.
They  are the parity eigenstates with eigenvalues of $\pm 1$. In
the parity operator the matrix $\gamma_0=\pmatrix{0&\openone\cr \openone &
0\cr}$ is used.

Let me turn now your attention to the helicity spin basis.
The 2-eigenspinors of the helicity operator 
\begin{eqnarray}
{1\over 2} {\bbox \sigma}\cdot\widehat
{\bf p} = {1\over 2} \pmatrix{\cos\theta & \sin\theta e^{-i\phi}\cr
\sin\theta e^{+i\phi} & - \cos\theta\cr}
\end{eqnarray}
can be defined as follows~\cite{Var,Dv1}:
\begin{eqnarray}
\phi_{{1\over 2}\uparrow}=\pmatrix{\cos{\theta \over 2} e^{-i\phi/2}\cr
\sin{\theta \over 2} e^{+i\phi/2}\cr}\,,\quad
\phi_{{1\over 2}\downarrow}=\pmatrix{\sin{\theta \over 2} e^{-i\phi/2}\cr
-\cos{\theta \over 2} e^{+i\phi/2}\cr}\,,\quad\label{ds}
\end{eqnarray}
for $\pm 1/2$ eigenvalues, respectively.

We start from the Klein-Gordon equation, generalized for
describing the spin-1/2  particles (i.~e., two degrees
of freedom); $c=\hbar=1$:
\begin{equation}
(E+{\bbox \sigma}\cdot {\bf p}) (E- {\bbox \sigma}\cdot {\bf p}) \phi
= m^2 \phi\,.\label{de}
\end{equation}
It can be re-written in the form of the set of two first-order equations
for 2-spinors. Simultaneously, we observe that they may be chosen
as eigenstates of the helicity operator which present
in (\ref{de}):\footnote{This opposes to the choice of the basis
(\ref{sb1}), where 4-spinors are the eigenstates of the parity operator.}
\begin{mathletters}
\begin{eqnarray}
(E-({\bbox\sigma}\cdot {\bf p})) \phi_\uparrow &=& (E-p) \phi_\uparrow
=m\chi_\uparrow \,,\\
(E+({\bbox\sigma}\cdot {\bf p})) \chi_\uparrow &=& (E+p) \chi_\uparrow
=m\phi_\uparrow \,,\\
(E-({\bbox\sigma}\cdot {\bf p})) \phi_\downarrow &=& (E+p) \phi_\downarrow
=m\chi_\downarrow\,, \\
(E+({\bbox\sigma}\cdot {\bf p})) \chi_\downarrow &=& (E-p) \chi_\downarrow
=m\phi_\downarrow \,.
\end{eqnarray}
\end{mathletters}
If the $\phi$ spinors are defined by the equation (\ref{ds}) then we
can construct the corresponding $u-$ and $v-$ 4-spinors\footnote{One can
also try to construct yet another theory differing from
the ordinary Dirac theory. The 4-spinors might be {\it not}
the eigenspinors of the helicity operator of the $(1/2,0)\oplus (0,1/2)$
representation space, cf.~\cite{DasG}. They might be
the eigenstates of the {\it chiral} helicity operator introduced
in~[2a]. In this case, the momentum-space Dirac equations
can be written (cf.~[2c],\cite{Dv2})
\begin{mathletters}
\begin{eqnarray}
p_\mu \gamma^\mu {\cal U}_\uparrow - m {\cal U}_\downarrow &=&0 \,,\\
p_\mu \gamma^\mu {\cal U}_\downarrow - m {\cal U}_\uparrow &=&0 \,,\\
p_\mu \gamma^\mu {\cal V}_\uparrow + m {\cal V}_\downarrow &=&0 \,,\\
p_\mu \gamma^\mu {\cal V}_\downarrow + m {\cal V}_\uparrow &=&0 \,.
\end{eqnarray}
\end{mathletters}
Here $\uparrow\downarrow$ refers already to the chiral helicity
eigenstates, e.g. $u_\eta ={1\over \sqrt{2}} \pmatrix{N \phi_\eta\cr
N^{-1} \phi_{-\eta}\cr}$.}
\begin{mathletters} \begin{eqnarray}
u_\uparrow &=&
N_\uparrow^+ \pmatrix{\phi_\uparrow\cr {E-p\over m}\phi_\uparrow\cr} =
{1\over \sqrt{2}}\pmatrix{\sqrt{{E+p\over m}} \phi_\uparrow\cr
\sqrt{{m\over E+p}} \phi_\uparrow\cr}\,,
u_\downarrow = N_\downarrow^+ \pmatrix{\phi_\downarrow\cr
{E+p\over m}\phi_\downarrow\cr} = {1\over
\sqrt{2}}\pmatrix{\sqrt{{m\over E+p}} \phi_\downarrow\cr \sqrt{{E+p\over
m}} \phi_\downarrow\cr}\,,\label{s1}\\
v_\uparrow &=& N_\uparrow^- \pmatrix{\phi_\uparrow\cr
-{E-p\over m}\phi_\uparrow\cr} = {1\over \sqrt{2}}\pmatrix{\sqrt{{E+p\over
m}} \phi_\uparrow\cr
-\sqrt{{m\over E+p}} \phi_\uparrow\cr}\,,
v_\downarrow = N_\downarrow^- \pmatrix{\phi_\downarrow\cr
-{E+p\over m}\phi_\downarrow\cr} = {1\over
\sqrt{2}}\pmatrix{\sqrt{{m\over E+p}} \phi_\downarrow\cr -\sqrt{{E+p\over
m}} \phi_\downarrow\cr}\,,\label{s2}
\end{eqnarray} \end{mathletters}
where the normalization to the unit
($\pm 1$) was used:\footnote{Of course, there are no any mathematical
difficulties
to change it  to the normalization to $\pm m$, which may be more convenient
for our study of the massless limit.}
\begin{mathletters}
\begin{eqnarray}
\bar u_\lambda u_{\lambda^\prime} &=& \delta_{\lambda\lambda^\prime}\,,
\bar v_\lambda v_{\lambda^\prime} = -\delta_{\lambda\lambda^\prime}\,,\\
\bar u_\lambda v_{\lambda^\prime} &=& 0 =
\bar v_\lambda u_{\lambda^\prime}
\end{eqnarray}
\end{mathletters}
One can prove that the matrix
\begin{equation}
P=\gamma^0 = \pmatrix{0&\openone\cr\openone & 0\cr}
\label{par}
\end{equation}
can be used in the parity operator as well as
in the original Dirac basis. Indeed, the 4-spinors
(\ref{s1},\ref{s2}) satisfy the Dirac equation in the spinorial
representation of the $\gamma$-matrices (see straightforwardly
from (\ref{de})). Hence, the parity-transformed function
$\Psi^\prime (t, -{\bf x})=P\Psi (t,{\bf x})$ must satisfy
\begin{equation}
[i\gamma^\mu \partial_\mu^{\,\prime} -m ] \Psi^\prime (t,-{\bf x}) =0 \,,
\end{equation}
with $\partial_\mu^{\,\prime} = (\partial/\partial t, -{\bf \nabla}_i)$.
This is possible when $P^{-1}\gamma^0 P = \gamma^0$ and
$P^{-1} \gamma^i P = -\gamma^i$. The matrix (\ref{par})
satisfies these requirements, as in the textbook case.

Next, it is easy to prove that one can form the projection
operators
\begin{mathletters}
\begin{eqnarray}
P_+ &=&+\sum_{\lambda} u_\lambda ({\bf p}) \bar u_\lambda ({\bf p})
=\frac{p_\mu \gamma^\mu +m}{2m}\,,\\
P_- &=&-\sum_{\lambda} v_\lambda ({\bf p}) \bar v_\lambda ({\bf p})
= \frac{m- p_\mu \gamma^\mu}{2m}\,,
\end{eqnarray}
\end{mathletters}
with the properties $P_+ +P_- =1$ and $P_\pm^2 = P_\pm$.
This permits us to expand the 4-spinors defined in the basis  (\ref{sb1})
in  linear superpositions of the helicity basis
4-spinors and to find corresponding coefficients of the expansion:
\begin{mathletters}
\begin{eqnarray}
u_\sigma ({\bf p}) &=& A_{\sigma\lambda} u_\lambda ({\bf p})
+ B_{\sigma\lambda} v_\lambda ({\bf p})\,,\\
v_\sigma ({\bf p}) &=& C_{\sigma\lambda} u_\lambda ({\bf p})
+ D_{\sigma\lambda} v_\lambda ({\bf p})\,.
\end{eqnarray}
\end{mathletters}
Multiplying the above equations by $\bar u_{\lambda^\prime}$,
$\bar v_{\lambda^\prime}$ and using the normalization conditions,
we obtain $A_{\sigma\lambda}= D_{\sigma\lambda}= \bar u_\lambda u_\sigma$,
$B_{\sigma\lambda}=C_{\sigma\lambda}= - \bar v_\lambda u_\sigma$.
Thus,  the transformation matrix from the common-used basis to the helicity basis is
\begin{equation}
\pmatrix{u_\sigma\cr
v_\sigma\cr}={\cal U} \pmatrix{u_\lambda\cr
v_\lambda\cr},\,\,\quad{\cal U} = \pmatrix{A&B\cr
B&A}
\end{equation}
Neither $A$ nor $B$ are unitary:
\begin{mathletters}
\begin{eqnarray}
A= (a_{++} +a_{+-}) (\sigma_\mu a^\mu) +(-a_{-+} +a_{--})
(\sigma_\mu a^\mu) \sigma_3\,,\\
B= (-a_{++} +a_{+-}) (\sigma_\mu a^\mu) +(a_{-+} +a_{--})
(\sigma_\mu a^\mu) \sigma_3\,,
\end{eqnarray}
\end{mathletters}
where
\begin{mathletters}
\begin{eqnarray}
a^0 &=& -i\cos (\theta/2) \sin (\phi/2) \in \Im m\,,\quad
a^1 = \sin (\theta/2) \cos (\phi/2)\in \Re e\,,\\
a^2 &=& \sin (\theta/2) \sin (\phi/2) \in \Re e\,,\quad
a^3 = \cos (\theta/2) \cos (\phi/2)\in \Re e\,,
\end{eqnarray}
\end{mathletters}
and
\begin{mathletters}
\begin{eqnarray}
a_{++} &=&\frac{\sqrt{(E+m)(E+p)}}{2\sqrt{2} m}\,,\quad
a_{+-} =\frac{\sqrt{(E+m)(E-p)}}{2\sqrt{2} m}\,,\\
a_{-+} &=&\frac{\sqrt{(E-m)(E+p)}}{2\sqrt{2} m}\,,\quad
a_{--} =\frac{\sqrt{(E-m)(E-p)}}{2\sqrt{2} m}\,.
\end{eqnarray}
\end{mathletters}
However, $A^\dagger A+B^\dagger B =\openone$, so the matrix ${\cal U}$
is unitary. Please note that this matrix acts
on the {\it spin}  indices ($\sigma$,$\lambda$), and not on
the spinorial indices; it is $4\times 4$ matrix. Alternatively,
the transformation can be written:
\begin{mathletters}
\begin{eqnarray}
u_\sigma^\alpha &=& [A_{\sigma\lambda}\otimes I_{\alpha\beta}
+B_{\sigma\lambda}\otimes \gamma^5_{\alpha\beta}] u_\lambda^\beta\,,\\
v_\sigma^\alpha &=& [A_{\sigma\lambda}\otimes I_{\alpha\beta}
+B_{\sigma\lambda}\otimes \gamma^5_{\alpha\beta}] v_\lambda^\beta\,.
\end{eqnarray}
\end{mathletters}

We now investigate the properties of the helicity-basis 4-spinors
with respect to the discrete symmetry operations $P,C$ and $T$.
It is expected that $\lambda\rightarrow -\lambda$ under parity,
as Berestetski\u{\i}, Lifshitz and Pitaevski\u{\i}
claimed~\cite{Lan}.\footnote{Indeed, if ${\bf x}\rightarrow -{\bf x}$,
then the vector ${\bf p}\rightarrow -{\bf p}$, but the axial vector
${\bf S}\rightarrow {\bf S}$, that implies the above statement.}
With respect to ${\bf p} \rightarrow -{\bf p}$ (i.~e.,
the spherical system  angles $\theta \rightarrow \pi-\theta$,
$\varphi \rightarrow \pi+\varphi$) the helicity 2-eigenspinors
transform as follows: $\phi_{\uparrow\downarrow} \Rightarrow
-i \phi_{\downarrow\uparrow}$, ref.~\cite{Dv1}.
Hence,
\begin{mathletters}
\begin{eqnarray}
Pu_\uparrow (-{\bf p}) &=& -i u_\downarrow ({\bf p})\,,
Pv_\uparrow (-{\bf p}) = +i v_\downarrow ({\bf p})\,,\\
Pu_\downarrow (-{\bf p}) &=& -i u_\uparrow ({\bf p})\,,
Pv_\downarrow (-{\bf p}) = +i v_\uparrow ({\bf p})\,.
\end{eqnarray}
\end{mathletters}
Thus, on the level of classical fields, we observe that
the helicity 4-spinors transform to the 4-spinors of the opposite
helicity.

Under the charge conjugation operation we have:
\begin{equation}
C =\pmatrix{0&\Theta\cr
-\Theta & 0\cr} {\cal K}\,.
\end{equation}
Hence, we observe
\begin{mathletters}
\begin{eqnarray}
Cu_\uparrow ({\bf p}) &=& - v_\downarrow ({\bf p})\,,
Cv_\uparrow ({\bf p}) = +  u_\downarrow ({\bf p})\,,\\
Cu_\downarrow ({\bf p}) &=& + v_\uparrow ({\bf p})\,,
Cv_\downarrow ({\bf p}) = - u_\uparrow ({\bf p})\,.
\end{eqnarray}
\end{mathletters}
due to the properties of the Wigner operator $\Theta \phi_\uparrow^\ast =
-\phi_\downarrow$ and $\Theta \phi_\downarrow^\ast = +\phi_\uparrow$.
For the $CP$ (and $PC$) operation we get:
\begin{mathletters}
\begin{eqnarray}
C P u_\uparrow (-{\bf p}) &=& -PC u_\uparrow (-{\bf p})
= +i v_\uparrow ({\bf p})\,,\\
C P u_\downarrow
(- {\bf p}) &=& - P C u_\downarrow (-{\bf p}) = -i v_\downarrow ({\bf
p})\,,\\
C P v_\uparrow (-{\bf p}) &=& - P C v_\uparrow (-{\bf p}) =
+  i u_\uparrow ({\bf p})\,,\\
C P v_\downarrow (-{\bf p}) &=& - P C v_\downarrow (-{\bf p}) =
- i u_\downarrow ({\bf p})\,.
\end{eqnarray}
\end{mathletters}
Similar conclusions can be drawn in the Fock space.
We define the field operator as follows:
\begin{equation}
\Psi (x^\mu) = \sum_\lambda \int \frac{d^3 {\bf p}}{(2\pi)^3}
\frac{\sqrt{m}}{2E} [ u_\lambda a_\lambda e^{-ip_\mu x^\mu} +v_\lambda
b^\dagger_\lambda e^{+ip_\mu x^\mu} ]\,.
\end{equation}
The commutation
relations are assumed to be the standard
ones~\cite{Bogol,Wein,Itzyk,Greib}\footnote{The only possible changes
may be related to a different form of normalization of 4-spinors,
which would have influence on the factor before $\delta$-function.}
(compare  with~\cite{DasG,Dv2})
\begin{mathletters}
\begin{eqnarray}
\left [a_\lambda ({\bf p}),
a_{\lambda^\prime}^\dagger ({\bf k})
\right ]_+ &=& 2E \delta^{(3)} ({\bf p}-{\bf k})
\delta_{\lambda\lambda^\prime}\,,
\left [a_\lambda ({\bf p}),
a_{\lambda^\prime} ({\bf k})\right ]_+ = 0 =
\left [a_\lambda^\dagger ({\bf p}),
a_{\lambda^\prime}^\dagger ({\bf k})\right ]_+\,,\\
\left [a_\lambda ({\bf p}),
b_{\lambda^\prime}^\dagger ({\bf k})\right ]_+ &=& 0 =
\left [b_\lambda ({\bf p}),
a_{\lambda^\prime}^\dagger ({\bf k})\right ]_+\,,\\
\left [b_\lambda ({\bf p}),
b_{\lambda^\prime}^\dagger ({\bf k})
\right ]_+ &=& 2E \delta^{(3)} ({\bf p}-{\bf k})
\delta_{\lambda\lambda^\prime}\,,
\left [b_\lambda ({\bf p}),
b_{\lambda^\prime} ({\bf k})\right ]_+ = 0 =
\left [b_\lambda^\dagger ({\bf p}),
b_{\lambda^\prime}^\dagger ({\bf k})\right ]_+\,.
\end{eqnarray} \end{mathletters}
If one defines $U_P \Psi (x^\mu) U_P^{-1} = \gamma^0  \Psi
(x^{\mu^\prime})$, $U_C \Psi (x^\mu) U_C^{-1} = \tilde C \Psi^\dagger (x^\mu)$
and the anti-unitary operator of time reversal $(V_T \Psi (x^\mu)
V_T^{-1})^\dagger = T \Psi^\dagger (x^{\mu^{\prime\prime}})$,
then it is easy to obtain the corresponding transformations
of the creation/annihilation operators (cf. the cited textbooks).
\begin{mathletters}
\begin{eqnarray}
U_P a_\lambda U_P^{-1} &=& -i a_{-\lambda} (-{\bf p})\,,
U_P b_\lambda U_P^{-1} = -i b_{-\lambda} (-{\bf p})\,,\label{pa1}\\
U_C a_\lambda U_C^{-1} &=& (-1)^{{1\over 2}+\lambda} b_{-\lambda} ({\bf
p})\,, U_C b_\lambda U_C^{-1} = (-1)^{{1\over 2}-\lambda} a_{-\lambda}
(-{\bf p})\,.
\end{eqnarray}
\end{mathletters}
As a consequence, we obtain (provided that $U_P \vert 0>=\vert 0>$,
$U_C\vert 0>= \vert 0>$)
\begin{mathletters}\begin{eqnarray}
&&U_P a^\dagger_\lambda ({\bf p}) \vert 0>=
U_P a_\lambda^\dagger U_P^{-1} \vert 0> =i a_{-\lambda}^\dagger
(-{\bf p}) \vert 0>= i \vert -{\bf p}, -\lambda >^+\,,\\ &&U_P
b^\dagger_\lambda ({\bf p}) \vert 0> =U_P b_\lambda^\dagger U_P^{-1}
\vert 0> =i b_{-\lambda}^\dagger (-{\bf p}) \vert 0>= i \vert -{\bf p},
-\lambda >^-\,;
\end{eqnarray}\end{mathletters}
and
\begin{mathletters}\begin{eqnarray}
&&U_C a^\dagger_\lambda ({\bf p}) \vert 0>=
U_C a_\lambda^\dagger U_C^{-1} \vert 0> = (-1)^{{1\over 2} +\lambda}
b_{-\lambda}^\dagger ({\bf p}) \vert 0>= (-1)^{{1\over 2}+\lambda} \vert
{\bf p}, -\lambda >^-\,,\\
&&U_C b^\dagger_\lambda ({\bf p}) \vert 0>= U_C
b_\lambda^\dagger U_C^{-1} \vert 0> = (-1)^{{1\over
2}-\lambda} a_{-\lambda}^\dagger ({\bf p}) \vert 0>= (-1)^{{1\over
2}-\lambda} \vert {\bf p}, -\lambda >^+\,.
\end{eqnarray}\end{mathletters}
Finally, for the $CP$ operation one should obtain:
\begin{mathletters}
\begin{eqnarray}
&&U_P U_C a^\dagger_\lambda ({\bf p}) \vert 0>=
-U_C U_P a^\dagger_\lambda ({\bf p}) \vert 0> = (-1)^{{1\over 2}+\lambda}
U_P b_{-\lambda}^\dagger ({\bf p}) \vert 0> =\nonumber\\
&=& i (-1)^{{1\over 2} +\lambda}
b_{\lambda}^\dagger (-{\bf p}) \vert 0>= i (-1)^{{1\over 2}+\lambda} \vert
-{\bf p}, \lambda >^-\,,\\
&&U_P U_C b^\dagger_\lambda ({\bf p}) \vert 0>= -U_C U_P b^\dagger_\lambda
({\bf p}) = (-1)^{{1\over 2}-\lambda} U_P
a_{-\lambda}^\dagger ({\bf p})\vert 0> = \nonumber\\
&=&i (-1)^{{1\over
2}-\lambda} a_{\lambda}^\dagger (-{\bf p}) \vert 0>= i (-1)^{{1\over
2}-\lambda} \vert -{\bf p}, -\lambda >^+\,.
\end{eqnarray}
\end{mathletters}
As in the classical case, the $P$ and $C$ operations anticommutes
in the $({1\over 2},0)\oplus (0,{1\over 2})$ quantized case. This opposes
to the theory based on 4-spinor eigenstates of chiral helicity
(cf.~\cite{Dv2}).

Since the $V_T$ is an anti-unitary operator the problem must be solved
after taking into account that in this case
the $c$-numbers should be put outside
the hermitian conjugation
{\it without} complex conjugation:
\begin{equation}
[V_T \lambda A V_T^{-1}]^\dagger = [\lambda^\ast V_T A V_T^{-1} ]^\dagger
= \lambda [V_T A^\dagger V_T^{-1} ]\,.
\end{equation}
With this definition we obtain:\footnote{$T$ is chosen to be
$T=\pmatrix{\Theta &0\cr
0& \Theta\cr}$ in order to fulfill
$T^{-1} \gamma_0^T T= \gamma_0$, $T^{-1} \gamma_i^T T= \gamma_i$
and $T^T= -T$.}
\begin{mathletters}
\begin{eqnarray}
V_T a_\lambda^\dagger V_T^{-1} &=& +i
(-1)^{{1\over 2}-\lambda}
a_{\lambda}^\dagger (-{\bf p})\,,\\
V_T b_\lambda  V_T^{-1} &=& +i (-1)^{{1\over 2}-\lambda}
b_{\lambda} (-{\bf p})\,.
\end{eqnarray}
\end{mathletters}

Furthermore, we observed that the question of
whether a particle and an antiparticle have the same or
opposite parities depend on a phase factor
in the following definition:
\begin{equation}
U_P \Psi (t, {\bf x}) U_P^{-1} = e^{i\alpha} \gamma^0 \Psi (t, -{\bf
x})\,.  \label{def1}
\end{equation}
Indeed, if we repeat the textbook procedure~\cite{Greib}:
\begin{eqnarray}
\lefteqn{U_P \left [ \sum_{\lambda}^{}
\int {d^3 {\bf p} \over (2\pi)^3} {\sqrt{m}\over 2E}
(u_\lambda ({\bf p}) a_\lambda ({\bf p}) e^{-ip_\mu x^\mu}
+v_\lambda ({\bf p}) b_\lambda^\dagger ({\bf p}) e^{+ip_\mu x^\mu}) \right
] U_P^{-1} =\,\nonumber}\\
&=& e^{i\alpha} \left [ \sum_{\lambda}^{} \int {d^3 {\bf
p} \over (2\pi)^3} {\sqrt{m}\over 2E} (\gamma^0 u_\lambda (-{\bf
p}) a_\lambda (-{\bf p}) e^{-ip_\mu x^\mu} +\gamma^0 v_\lambda
(-{\bf p}) b_\lambda^\dagger (-{\bf p}) e^{+ip_\mu x^\mu}) \right ]= \,
\nonumber\\
&=& e^{i\alpha} \left [ \sum_{\lambda}^{} \int {d^3 {\bf
p} \over (2\pi)^3} {\sqrt{m}\over 2E} (-i u_{-\lambda} ({\bf
p}) a_\lambda (-{\bf p}) e^{-ip_\mu x^\mu} +i v_{-\lambda}
({\bf p}) b_\lambda^\dagger (-{\bf p}) e^{+ip_\mu x^\mu}) \right ]\,.
\end{eqnarray}
Multiplying by $u_{\lambda^\prime} ({\bf p})$
and $v_{\lambda^\prime} ({\bf p})$ consequetively,
and using the normalization conditions we obtain
\begin{mathletters}
\begin{eqnarray}
&&U_P a_\lambda U_P^{-1} = -i e^{i\alpha} a_{-\lambda} (-{\bf p})\,,\\
&&U_P b_\lambda^\dagger U_P^{-1} = + i e^{i\alpha} b_{-\lambda}^\dagger
(-{\bf p})\,.
\end{eqnarray}
\end{mathletters}
From this, if $\alpha=\pi/2$ we obtain
{\it opposite} parity properties of creation/annihilation
operators for particles and anti-particles:
\begin{mathletters}
\begin{eqnarray}
&&U_P a_\lambda U_P^{-1} = + a_{-\lambda} (-{\bf p})\,,\\
&&U_P b_\lambda U_P^{-1} = - b_{-\lambda}
(-{\bf p})\,.
\end{eqnarray}
\end{mathletters}
However, the difference with the Dirac case still preserves
($\lambda$ transforms to $-\lambda$). As a conclusion,
the question of the same (opposite) relative intrinsic parity
is intrinsically related to the phase factor in (\ref{def1}).
We find somewhat similar situation with the  question of constructing
the neutrino field operator (cf. with the Goldhaber-Kayser
creation phase factor).

Next, we find the explicit form of the parity operator
$U_P$ and prove that it commutes with the Hamiltonian operator.
We prefer to use the method described in~\cite[\S 10.2-10.3]{Greib}.
It is based on the anzatz that $U_P = \exp [i\alpha \hat A] \exp [i \hat
B]$ with $\hat A =\sum_{s}^{}\int d^3 {\bf p} [a_{{\bf p},s}^\dagger a_{-{\bf
p}s} +b_{{\bf p}s}^\dagger b_{-{\bf p}s}]$ and $\hat B =\sum^{}_{s}\int d^3
{\bf p} [\beta a_{{\bf p},s}^\dagger a_{{\bf p}s} +\gamma b_{{\bf
p}s}^\dagger b_{{\bf p}s}]$. On using the known operator identity
\begin{equation}
e^{\hat A} \hat B e^{-\hat A} = \hat B +[\hat A,\hat B]_- +{1\over 2!}
[\hat A, [\hat A,\hat B]]+\ldots
\end{equation}
and $[\hat A,\hat B\hat C]_-= [\hat A,\hat B]_+ \hat C
-\hat B [\hat A,\hat C]_+$ one can fix  the parameters
$\alpha,\beta,\gamma$ such that satisfy the physical
requirements that a Dirac particle and its anti-particle
have opposite intrinsic parities.

In our case, we  need to satisfy (\ref{pa1}), i.e., the operator
should invert not only the sign of the momentum, but the sign of
the helicity too. We may achieve this goal by the analogous postulate
$U_P= e^{i\alpha \hat A}$ with
\begin{equation}
\hat A =\sum_{s}^{} \int {d^3 {\bf p}\over 2E}
[a^\dagger_\lambda ({\bf p}) a_{-\lambda} (-{\bf p})
+b_\lambda^\dagger ({\bf p}) b_{-\lambda} (-{\bf p})]\,.
\end{equation}
By direct verification, the equations (\ref{pa1})
are satisfied provided that $\alpha=\pi/2$. Cf. this parity operator with
that given in~\cite{Itzyk,Greib} for Dirac fields:\footnote{Greiner used the following commutation
relations
$\left [ a ({\bf p}, s), a^\dagger ({\bf p}^\prime, s^\prime) \right ]_+ =
\left [ b ({\bf p}, s), b^\dagger ({\bf p}^\prime, s^\prime) \right ]_+ =
\delta^3 ({\bf p}-{\bf p}^\prime) \delta_{ss^\prime}$. One should also
note that the Greiner form of the parity operator is not the only one.
Itzykson and Zuber~\cite{Itzyk} proposed another one differing by the
phase factors from (10.69) of~\cite{Greib}.
In order to find relations between those
two forms of the parity operator one should apply
additional rotation in the Fock space.}
\begin{eqnarray} \lefteqn{U_P =
\exp \left [ i{\pi\over 2} \int  d^3 {\bf p} \sum_s
\left ( a ({\bf p}, s)^\dagger
a (\tilde{\bf p},s) +b ({\bf p},s)^\dagger b (\tilde{\bf p},s)-
\right.\right.}\nonumber\\
&&\left.\left.- a ({\bf p},s)^\dagger a ({\bf p},s) + d ({\bf
p},s)^\dagger b ({\bf p},s) \right ) \right ]\,,\quad (10.69)\,\,
\mbox{of}\,\,\,
\mbox{ref.}~\cite{Greib}\,.\end{eqnarray}
By
direct verification one can also come to the conclusion that  our new $U_P$ commutes
with the Hamiltonian:  \begin{equation} {\cal H} = \int d^3 {\bf x}
\Theta^{00} = \int d^3 {\bf k} \sum_\lambda [ a_\lambda^\dagger ({\bf k})
a_\lambda ({\bf k}) - b_\lambda ({\bf k}) b_\lambda^\dagger ({\bf k})]\,,
\end{equation} i.e.
\begin{equation}
[U_P, {\cal H} ]_- =0\,.
\end{equation}
Alternatively, we can try to choose another set of commutation
relations~[2b,3] (for the set
of bi-orthonormal states), that will be the matter of
future publications.

Finally, due to the fact that my recent works are related to the
so-called ``Bargmann-Wightman-Wigner-type" quantum field theory,
I want to clarify some misunderstandings in the recent discussions.
This type of theories has been first proposed by Gel'fand and
Tsetlin~[19a]. In fact, it is based on the two-dimensional representation
of the inversion group, which is used when someone needs to construct a theory
where $C$ and $P$ anticommute. They
indicated applicability of this theory to the description of the set of
$K$-mesons and possible relations to the Lee-Yang result.
The comutativity/anticommutativity of the discrete symmetry operations has
also been investigated by Foldy and Nigam~\cite{FN}. Relations of
the Gel'fand-Tsetlin construct to the representations of the
anti-de Sitter $SO(3,2)$ group and the general relativity theory
(including continuous and discrete transformations) have been discussed
in~[19b] and in subsequent papers of Sokolik. E. Wigner~\cite{Wig}
presented somewhat related results at the Istanbul School
on Theoretical Physics in 1962. Later, Fushchich discussed
corresponding wave equations. At last, in the paper~\cite{Dva}
the authors called a theory where a boson and its antiboson
have opposite intrinsic parities as the theory of ``the
Bargmann-Wightman-Wigner type". Actually, the theory presented
by Ahluwalia, Goldman and Johnson is
the Dirac-like generalization of the Weinberg $2(2J+1)$-theory
for the spin 1. It has already been presented in the Sankaranarayanan
and Good paper of 1965, ref.~\cite{SG}. In ref.~[22b] (and
in the previous IF-UNAM preprints of 1994) I presented a theory
based on a set of 6-component Weinberg-like equations (I called
them the ``Weinberg doubles"). In ref.~[2b] the theory in the $({1\over
2},0)\oplus (0,{1\over 2})$ representation based on
the chiral helicity 4-eigenspinors was proposed. The connection with the Foldy
and Nigam consideration has been claimed. The corresponding equations
have been obtained in~\cite{Dv2}  and in several less known articles.
However, later we found the papers by Ziino and Barut~\cite{Barut}
and the Markov papers~\cite{Markov}, which also have connections
with the subject under consideration.

A similar theory may be constructed from our consideration above
if we define the field operators as follows:
\begin{mathletters}
\begin{eqnarray}
\Psi_1 &=&
\int {d^3 {\bf p}\over (2\pi)^3}
{\sqrt{m}\over 2E} \left [(u_\uparrow a_\uparrow
+v_\uparrow b_\uparrow) e^{-ip_\mu x^\mu}
+ (u_\uparrow a_\uparrow^\dagger
+v_\uparrow b_\uparrow^\dagger) e^{+ip_\mu x^\mu}\right ]\,,\\
\Psi_2 &=&
\int {d^3 {\bf p}\over (2\pi)^3}
{\sqrt{m}\over 2E} \left [(u_\downarrow a_\downarrow
-v_\downarrow b_\downarrow) e^{-ip_\mu x^\mu}
+ (u_\downarrow a_\downarrow^\dagger
-v_\downarrow b_\downarrow^\dagger) e^{+ip_\mu x^\mu}\right ]\,.
\end{eqnarray}
\end{mathletters}

The conclusions of my talk are:

\begin{itemize}

\item
Similarly to the $({1\over 2},{1\over 2})$ representation,
the $({1\over 2},0)\oplus (0,{1\over 2})$ field functions
in the helicity basis are {\it not} eigenstates of the
common-used parity operator; $\vert {\bf p},\lambda> \Rightarrow
\vert -{\bf p},-\lambda >$  both on the classical and quantum levels. This
is in accordance with the earlier consideration of Berestetski\u{\i},
Lifshitz and Pitaevski\u{\i}.

\item
Helicity field functions may satisfy the ordinary Dirac equation
with $\gamma$'s to be in the spinorial representation. Meanwhile, the chiral
helicity field functions satisfy the equations
of the form $\hat p \Psi_1 - m\Psi_2 =0$.

\item
Helicity field functions can be expanded in the set of the Dirac
4-spinors by means of the matrix ${\cal U}^{-1}$ given in this paper.
Neither $A$, nor $B$ are unitary, however $A^\dagger A+B^\dagger B =
\openone$.

\item
$P$ and $C$ operations anticommute  in this framework, both on the
classical and quantum levels (this is opposite to the theory based on the
chiral helicity eigenstates~\cite{Dv2}.

\item
Particle and antiparticle may have either the same or the opposite
properties
with respect to parity. The answer depends on the choice of the phase
factor in
$U_P \Psi U_P^{-1} = e^{i\alpha} \gamma^0 \Psi^\prime$; alternatively,  that 
can be made by additional rotation $U_{P_2}$.

\item
Earlier confusions in the discussion of the
Gelfand-Tsetlin-Sokolik-Nigam-Foldy-Bargmann-Wightman-Wigner-type
(GTsS-NF-BWW) quantum field theory have been clarified.

\end{itemize}

\acknowledgments
I am grateful for discussions to Prof. Z. Oziewicz and participants
of the Clifford Algebras Conference and the Graph-Operads-Logic
Conference (May 2002).  

\nopagebreak

\end{document}